\documentclass[%
 reprint,
superscriptaddress,
preprintnumbers,
 amsmath,amssymb,
 aps,
]{revtex4-2}

\usepackage{setspace}
\AtBeginEnvironment{thebibliography}{\setstretch{0.9}}

\usepackage{lineno}

\usepackage{graphicx}
\usepackage{dcolumn}
\usepackage{bm}

\renewcommand{\v}[1]{\ensuremath{\mathbf{#1}}} 
 
\renewcommand\eqref[1]{Eq.\;\ref{#1}} 

\usepackage [english]{babel}
\usepackage [english = american]{csquotes}
\MakeOuterQuote{"}

\usepackage{color}					

\begin{document}

\title{The cell as a token: high-dimensional geometry in language models and cell embeddings}
\author{William Gilpin}
 \email{wgilpin@utexas.edu}
\affiliation{%
Department of Physics, The University of Texas at Austin, Austin, Texas 78712, USA
}%

\date{\today}

\begin{abstract}
Single-cell sequencing technology maps cells to a high-dimensional space encoding their internal activity.
Recently-proposed virtual cell models extend this concept, enriching cells' representations based on patterns learned from pretraining on vast cell atlases.
This review explores how advances in understanding the structure of natural language embeddings informs ongoing efforts to analyze single-cell datasets.
Both fields process unstructured data by partitioning datasets into tokens embedded within a high-dimensional vector space.
We discuss how the context of tokens influences the geometry of embedding space, and how low-dimensional manifolds shape this space's robustness and interpretation.
We highlight how new developments in foundation models for language, such as interpretability probes and in-context reasoning, can inform efforts to construct cell atlases and train virtual cell models.
\end{abstract}

\maketitle

\clearpage


Modern single-cell experiments \textit{decompile} the cell---abstracting it away from its squishy context, and rendering it as a single point in a high-dimensional vector space. Computational workflows attempt to invert this process: spatial transcriptomics recovers information about a cell's position, while lineage tracing reconstructs developmental stages. Recent efforts to construct virtual cells---massive machine learning models built upon language model architectures---represent the next step of this process. Trained on vast amounts of genomic data, these models aim to provide richer, more informative representations than the raw count matrices produced by single cell experiments.

How do we know if the representations learned by virtual cell models are meaningful?
If a single-cell embedding exactly matches known regulatory and developmental mechanisms, an embedding space may be accurate yet uninformative for making new discoveries. 
Conversely, if this space fails to recapitulate known relationships, it is difficult to attribute this to novelty or inaccuracy.
A similar problem arises in statistical learning. Large language models are trained on vast, unannotated volumes of text. To represent diverse text sources consistently, these models initially split input text into discrete tokens: minimal units consisting of words or word fragments.
They then convert these tokens into vectors in a high-dimensional space, a representation that enables further processing by modern, continuously-valued learning models.

The success of language models stems, in part, from the ability of language embeddings to accurately encode syntactic and semantic structure in high-dimensional spaces. The unique properties of high-dimensional geometry allow embeddings to effectively encode the semantic structure of language along low-dimensional manifolds, mirroring findings from single-cell biology in identifying developmental pathways and rare cell types.
What insights can single-cell embeddings gain from the statistical learning community? Here, we review recent developments and commonalities between these fields, highlighting general principles of language embeddings that may inform ongoing work in single-cell genomics.

\subsection{Context shapes the geometry of embeddings}

\begin{figure}
{
\centering
\includegraphics[width=\linewidth]{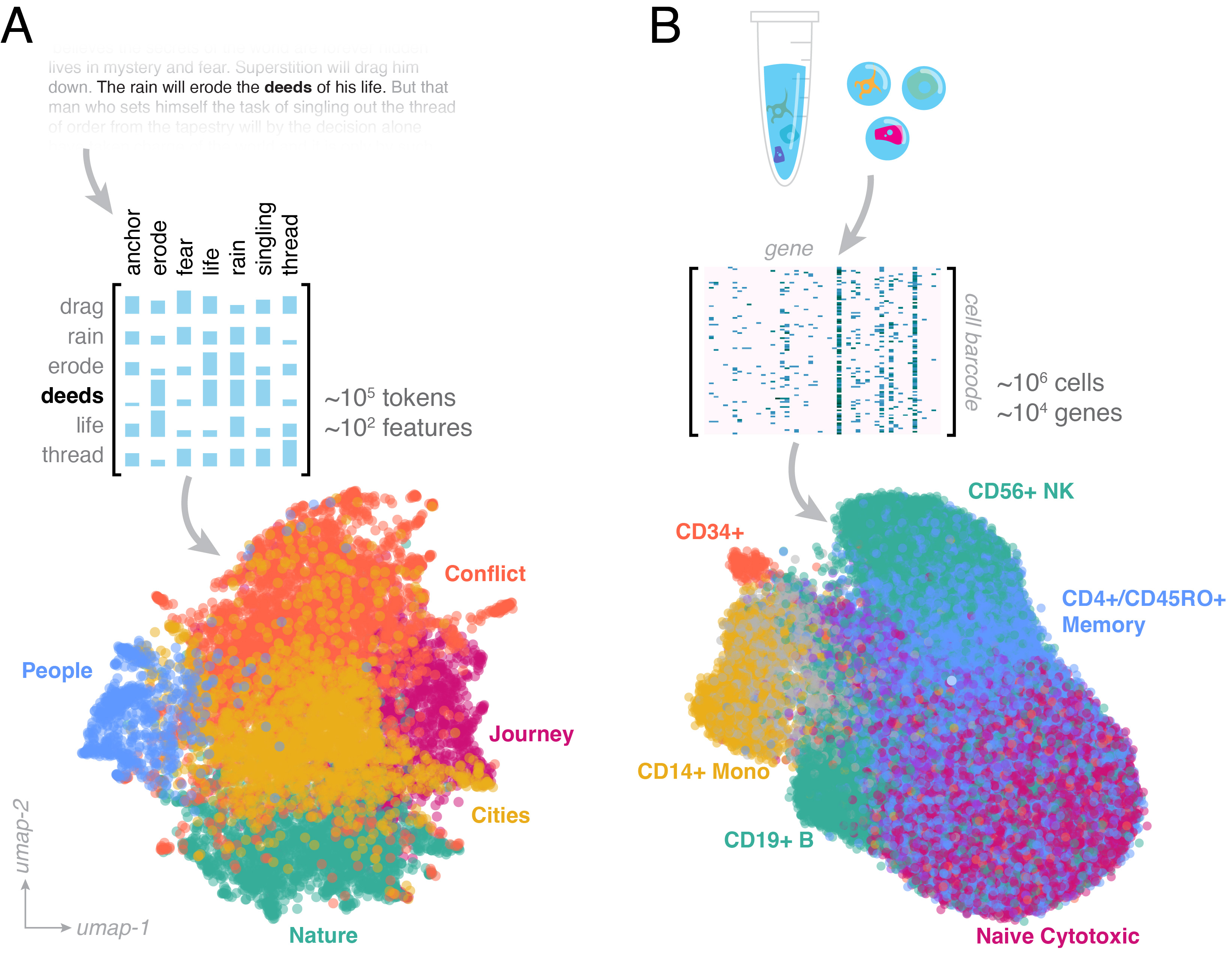}
\caption{
\textbf{Low-rank structure in high-dimensional embeddings.} 
(A) An embedding of the full text of the novel \textit{Blood Meridian} using a \texttt{word2vec} model originally trained on a dataset of $10^{11}$ words drawn from Google News articles \cite{mikolov2013distributed}. Vectors are clustered using K-means partitioning, and then summarized into metagroups with a topic embedding model (colors and annotations).
(B) An embedding of $6 \times 10^4$ human peripheral blood mononuclear cells based on single-cell RNA sequencing of $1.6 \times 10^4$ genes. Colors correspond to immune cell subtypes, as determined by marker genes for characteristic cell surface proteins like CD4, CD8, etc.
}
\label{fig:embeddings}
\vspace{-5mm}
}
\end{figure}

Modern large language models owe their scale to self-supervised training, which obviates the need to collect expensive labeled training data.  Given a sequence of words, "The parliamentarian led the assembly.", a single word is masked and the model is trained to fill in the blank: "The \texttt{[TOK]} led the assembly." When trained at scale, models learn to group certain words (e.g. "leader," "president," "speaker") that appear in similar contexts.
Theoretical motivation for context masking comes from the \textit{distributional hypothesis}, which equates distances between vector representations of different words in embedding space, with distances between distributions of co-occuring tokens within the training corpus \cite{harris1954distributional,firth1957synopsis}. 
The distributional hypothesis typically describes co-occurrence statistics: words like "president" or "parliamentarian" often appear with similar other words. However, the distribution may be further conditioned on language type, historical era, domain-specific register, or other latent variables that modulate word usage. In systems biology, the distributional hypothesis motivates efforts to train self-supervised foundation models from single cell data, often termed "virtual cells" \cite{bunne2024build}. An implicit assumption of such approaches is that models can learn informative, predictive knowledge purely from training to be self-consistent. Such approaches inherently invoke a distributional hypothesis---that cells occurring in the same tissues, interactions, or regulatory roles ought to retain that similarity when represented in a single-cell workflow, and that this similarity can be exploited for self-supervised training.

Predating modern large language models, \texttt{word2vec} language embeddings introduced an early notion of \textit{context} to word representations. 
During training, \texttt{word2vec} directly optimizes an objective function motivated by the distributional hypothesis, producing an embedding that maximizes the posterior probability of word-context pairs seen in the corpus, while minimizing the probability of randomly-generated pairs \cite{mikolov2013efficient}. 
This approach represents contrastive learning, which allows an embedding space to be constructed for data that otherwise lacks a well-defined distance metric.
A typical fully-trained \texttt{word2vec} model maps each of $10^7$ distinct words to a point in a $300$-dimensional continuous vector space. 
Because the distributional objective is optimized only during training on the text corpus, \texttt{word2vec} produces \textit{static embeddings}: after training, any appearance of a given token always maps to the same point in embedding space. 

Cell gene expression profiles lack an an obvious distance metric, and the results of computational workflows like cell type clustering vary depending on the choice of cell-cell distance metric such as Euclidean distance, correlation, or t-statistic \cite{ji2023optimal}. 
Raw expression profiles are typically context-independent. After isolation, sequencing, and demultiplexing, a cell becomes a collection of RNA transcripts, each of which may be considered a vector approximating the transcript counts per gene per cell \cite{stuart2019integrative}. 
The expression levels of each gene thus uniquely determine the embedding, decoupling a given cell's representation from others. 
Thus, in principle, raw count data do not invoke the distributional hypothesis: a cell's embedding is an innate property, rather than a property relative to a corpus of cells. 
Many preprocessing schemes applied to count matrices---such as batch or cell cycle correction---enforce static, context-free structure in embedding space \cite{korsunsky2019fast}.
However, data reduction methods like principal component analysis for visualization, or unsupervised clustering for cell type identification, produce context-dependent representations that depend on relative differences among cells. 
Context-dependence also arises when multiple datasets are merged, or when end-to-end embedding models are trained across many datasets. However, such approaches stop short of invoking the distributional hypothesis, because they do not enforce a notion of context tied to co-occurrence statistics. 
In contrast, the extensive pretraining used in modern single-cell foundation models aims to learn a distance metric among expression profiles based on statistical patterns in expression across the training data \cite{heimberg2025cell}.

\subsection{The geometry of embedding spaces} 

Theoretical analysis of \texttt{word2vec} and its variants shows that these methods, in practice, factorize a matrix representing the mutual information between the distribution of each token across the corpus, and the distribution of its context (Fig. \ref{fig:embeddings}A) \cite{levy2014neural}.
Linguistic structures, such as synonym clusters, lead to low-rank structure in this matrix \cite{dhillon2015eigenwords,allen2019vec}, similar to the low-rank structure that forms in single-cell count matrices due to statistical similarities in the expression vectors of cells belonging to the same type (Fig. \ref{fig:embeddings}B) \cite{nitzan2019gene}. 
Low-dimensional manifolds in single-cell embeddings typically arise from highly-coordinated biological processes, such as differentiation, which exhibit predominantly deterministic dynamics. In language embeddings, low-rank structure arises due to overparametrization---the highest-rank word-context matrix would simply represent an isotropic Gaussian distribution. Anisotropy in high-dimensional embeddings thus indicates structure in the underlying generative process, whether linguistic or biological.

A key limitation of static language embeddings stems from polysemy, in which the same token has multiple meanings \cite{gari2021let,liu2020survey}.
For example, "bank" may refer to the side of a riverbed, or to a financial institution. Static word embeddings like \texttt{word2vec} place polysemous tokens at intermediate positions in embedding space, between positions associated with their divergent meanings.
Such compromises distort and curl embedding space, reducing the space's ability to represent large-scale structure by making distances between vectors less meaningful \cite{jakubowski2020topology,neelakantan2014efficient,goel2022unsupervised}.
As a result, static embedding models tend to underestimate differences among strongly-distinct concepts, limiting their ability to recognize hierarchies among words \cite{nickel2017poincare}.
In gene expression, curvature due to polysemy may arise due to biological processes, rather than artifacts. Cellular differentiation datasets exhibit low curvature in regions associated with stereotyped cell states, punctuated by high-curvature regions associated with transitions \cite{sritharan2021computing}. These transition states, such as differentiating stem cells, occupy intermediate locations in embedding space \cite{wang2020single}. 
However, spurious polysemy can also arise due to technical errors, leading to unresolved cell subtypes or tissue groups. This effect becomes more pronounced at low read depth, resulting in missing genes or greater sampling error in counts.
For example, blood vascular endothelial cells share relatively consistent transcriptional profiles, due to their similar structural roles across different tissues \cite{kalucka2020single}. Endothelial cells from different tissues often map to the same area in embedding space, despite their anatomical separation. 
Resolving this ambiguity either requires additional marker genes and greater sequencing resolution, or assays that barcode transcripts with additional information. For example, cell painting produces a high-dimensional vector of morphological features extracted from fluorescence microscopy \cite{bray2016cell}, while CITE-Seq augments each transcript with information about cell surface proteins---thus avoiding cellular polysemy \cite{stoeckius2017simultaneous}.

Contemporary language models use dynamic token embeddings, in which a given token's embeddings varies based on its context after training \cite{devlin2019bert,liu2020survey,reisinger2010multi}.
The standard mechanism, self-attention, combines a token's static representation, neighboring context tokens, and a positional encoding \cite{vaswani2017attention}. The resulting joint representation thus varies even after training when it appears in new contexts. Thus, while static embeddings associate each token with a single embedding point, dynamic embeddings map each token to a cloud of points capturing the diverse contexts in which it appears. The distance between the same token in different contexts is smaller than the distance between tokens, consistent with low-dimensional, anisotropic structure \cite{ethayarajh2019contextual}. 

In large-scale gene expression datasets like cell atlases, dynamic cell embeddings improve the global structure of representations.
Spatial transcriptomics augments each transcript with information about the cell's absolute spatial position, or relative position among neighboring cells. As a result, embeddings learned by these methods encode an underlying metric, and so both local and global distances are meaningful \cite{tian2023expanding,nitzan2019gene}. More abstract context information, such as organ group or tissue annotations, improves embeddings by disambiguating similar transcriptional profiles arising in distinct contexts \cite{xu2021probabilistic}. 
Conceptually, these approaches resemble language models that combine tokenization with queries to an external database that provides richer context \cite{borgeaud2022improving,khandelwal2019generalization}. This can include structured sources of information about tokens, like encyclopedias or human-curated concept maps \cite{speer2017conceptnet,zhang2019ernie}. The resulting models capture global relationships among concepts, without requiring substantial additional training. 
For single-cell data, similar approaches enrich transcript information with tissue or preexisting cell type annotations \cite{brbic2020mars,lin2022scjoint,lotfollahi2022mapping}, gold-standard experimental associations or transcription factors \cite{lee2024chrombert}, gene ontologies \cite{yuan2024cell}, or even topic information from scientific literature databases \cite{zhao2021learning,istrate2024scgenept}.
Other efforts pair each cell with gene-level context (such as sequence position or chromatin accessibility) to highlight mechanistic relationships \cite{chen2024simba,fu2025foundation}.

When single cell foundation models are trained using self-supervision, their internal representations can be extracted and used as dynamic embeddings of expression vectors
\cite{fang2024scmae,song2021scgcn,eraslan2019single,fu2025foundation,zhao2021learning,istrate2024scgenept}. Multimodal models produce richer representations by training on both expression data and external information, such as known regulatory hierarchies \cite{zhao2021learning,istrate2024scgenept,bunne2024build,lopez2018deep}. Many such approaches use self-attention to dynamically process tokens, as well as contrastive learning, producing representations with an underlying similarity metric---representing a form of distributional hypothesis for expression vectors \cite{cui2024scgpt,theodoris2023transfer,vaswani2017attention,han2022self}.

\subsection{Are cells or genes the "words" of single-cell biology?} 

Many large-scale pretrained models for genomic data directly adapt language architectures, treating the genome as a large body of text, with nucleotides acting as an alphabet and genes as words \cite{consens2025transformers,ji2021dnabert,levine2024cell2sentence,rizvi2025scaling,rosen2023universal,cui2024scgpt}.
Genes thus may seem to be a more natural analogue to words in statistical learning frameworks.
However, this equivalence has limits: genes do not recur within a single genome, and so identifying variations in their function across cells or individuals requires expression information, a quantity without an obvious analogue in natural language.
Instead, from the perspective of the distributional hypothesis, cells, not genes, represent minimal tokens, because similarity among cells can be inferred from recurring patterns across different biological contexts. As a result, many proposed applications of virtual cell models, such as cell type identification or lineage tracing, implicitly treat cells as words \cite{bunne2024build,pearce2025cross}. 
Context thus arises from neighboring cells, tissue microenvironments, or developmental stages, while genes represent the fixed vocabulary describing each cell token. 
Ambiguities about the correct unit of tokens also exist in language models: while many language models treat words as tokens, others use finer-grained units such as characters \cite{boukkouri2020characterbert} or even raw byte sequences \cite{xue2022byt5}.
Similarly, in biological settings, the choice of "token" is not fixed \textit{a priori}, but should be defined at the level where meaningful context recurs.

\subsection{Analogies as manifolds in embedding space}

\begin{figure}
{
\centering
\includegraphics[width=\linewidth]{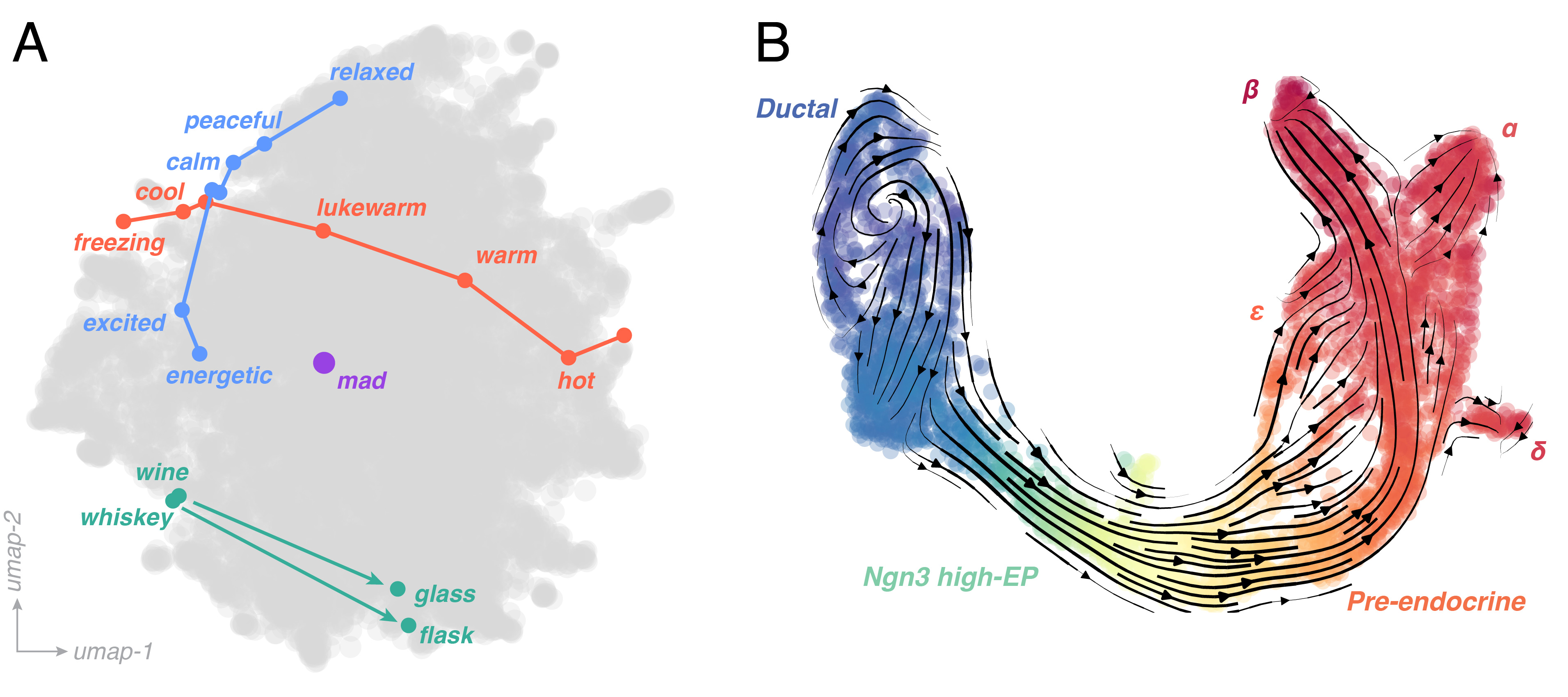}
\caption{
\textbf{Analogies and low-dimensional manifolds.} 
(A) Embeddings of particular sequences of tokens using the model of Fig. \ref{fig:embeddings}, with examples of escalating manifolds (red and blue lines), which overlap in regions with similar meaning (weak polysemy). A token with strong polysemy appears at an intermediate location (purple circle). An example of an analogy relationship encoded as nearly-congruent difference vectors (turquoise arrows). While nonlinear embedding methods like UMAP distort the local metric over large scales \cite{chari2023specious}, the nearby position of the two analogy vectors' heads and tails protects their congruency.
(B) RNA Velocity applied to developing endocrine cells in the pancreas \cite{bastidas2019comprehensive,la2018rna}. Vectors correspond to development direction, and color corresponds to pseudotime assigned via diffusion components. Cell types along the differentiation axis are overlaid.
}
\label{fig:analogies}
\vspace{-5mm}
}
\end{figure}

Effective language embeddings encode semantic relationships as distances \cite{mikolov2013linguistic}.
For example, the vector from "Sacramento" to "California" in embedding space may match the vector from "Austin" to "Texas." 
As a result, vector arithmetic in \texttt{word2vec} solves unseen analogy problems from college admissions exams, even without retraining (Fig. \ref{fig:analogies}A)\cite{liu2017analogical,turney2005corpus}.
Embedding space thus unfolds computation into a higher-dimensional space in which reasoning coincides with distances \cite{ushio2021bert}. 
In this sense, early word embeddings foreshadowed modern works on in-context learning and zero-shot inference, phenomena in which sufficiently-large models are able to perform new tasks not seen during training \cite{kojima2022large}.

Organismal cell atlases exhibit well-defined clusters associated with cell and tissue types. However, cells with different compositions but similar functions can nonetheless occupy similar relative locations in embedding space.
For example, immune cells form subtypes within different organ groups, such as Kupffer cells in the liver or microglia in the brain. In whole-organism cell atlases, these cells typically appear in separate clusters associated with their primary organ groups. However, within each organ cluster, they occupy similar positions relative to other cells, underscoring their analogous roles \cite{wang2022endothelial,suo2022mapping,gautier2012gene}. 

Language embeddings also capture continuous relationships among tokens. For example, escalating sequences like "good," "better," or "best" map to linear sequences in \texttt{word2vec} \cite{mikolov2013linguistic}. 
Generally, word embeddings exhibit high anisotropy \cite{mimno2017strange,ethayarajh2019contextual}, with embeddings spanning low-dimensional manifolds within the higher-dimensional representation space. Depending on the language, this manifold has effective dimensionality $\sim\!10^1$, even when the feature dimension is $\sim\!10^2$ \cite{mu2017all}.
These manifolds capture gradations in meanings among similar words, shifts in a word's meaning over time, singular-plural pairs, or even groups of synonyms \cite{hamilton2016cultural}.
For example, in dynamic embeddings produced by large language models, days of the week and calendar months map onto circular manifolds, while colors and years map onto linear manifolds \cite{engels2025not,modell2025origins}.
Consistent with these manifolds representing informative subspaces, the performance of embeddings in downstream tasks initially increases with the embedding dimension, but it eventually plateaus at a fixed multiple of the manifold dimension \cite{yin2018dimensionality}. 

Similar low-dimensional manifolds arise in cell embeddings. Across different datasets, cell replication cycles and circadian rhythms form rings \cite{kowalczyk2015single}, spatially-extended tissues form grids \cite{adler2019continuum,nitzan2019gene}, and cell differentiation hierarchies form branches (Fig. \ref{fig:analogies}B) \cite{paul2015transcriptional}.
In one well-known case, populations of blood cells of mixed maturity form a pitchfork in embedding space, illustrating a continuous progression from stem cells to different types of blood cells \cite{paul2015transcriptional}.
Gene expression manifolds have a typical intrinsic dimensionality $\sim 10^1$, compared to the $\sim 10^5$ genes measured in a typical single-cell experiment \cite{sritharan2021computing}. 
Just as word embeddings trace their properties to low-rank structure in the word-context mutual information matrix, biological processes confer low-rank structure on count matrices \cite{nitzan2021revealing,thibeault2024low}.

Theoretical models of word embeddings frame text generation as a stochastic dynamical system, with sentence formation as a random walk in token embedding space \cite{arora2016latent,arora2015random,hashimoto2016word}. Under this framework, semantic manifolds act as kinetic traps for the walk, with synonymous tokens acting as basins, and connective phrases acting as bridges. In single-cell analysis, diffusion maps simulate the action of many random walks through expression space \cite{coifman2006diffusion,wolf2019paga}. These methods represent a standard approach to calculating pseudotime, which orders unsorted cell embeddings to identify temporal progressions of cells (like developmental stages) \cite{haghverdi2016diffusion,setty2019characterization}.
In dynamic word embeddings, or in static embeddings trained on corpora from different historical periods, the relative positions of words gradually shift over time. This semantic drift may be quantified using a calculation resembling pseudotime \cite{bamler2017dynamic}.

\begin{figure*}
{
\centering
\includegraphics[width=\linewidth]{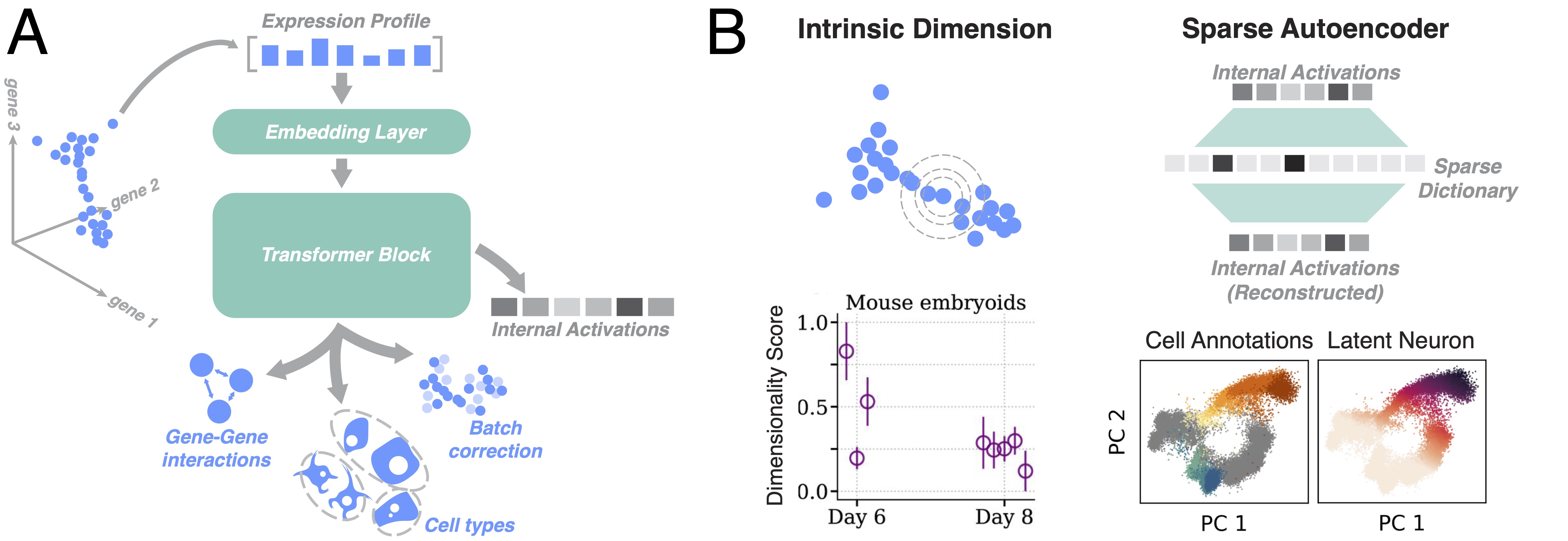}
\caption{
\textbf{Mechanistic interpretability in single-cell foundation models.} 
(A) Common architectural features and target tasks for single-cell foundation models.
(B) Mechanistic interpretability methods for single cell embeddings. (Left) Intrinsic dimensionality may be calculated directly from expression profiles, or from internal activations of the model. Inset shows the intrinsic dimensionality of staged expression profiles from developing mice. Panel adapted from Ref. \cite{biondo2024intrinsic}. (Right) Sparse autoencoders are trained in an unsupervised manner to reconstruct internal activations of foundation models, by mapping activations to sparse combinations of features in a latent dictionary. Inset shows application of sparse autoencoders to the activations of the Universal Cell Embedding model on a dataset of human bone marrow. The left subpanel corresponds to annotated cell types, while the right corresponds to the decoding of a single latent unit.
Panels adapted from Ref. \cite{schuster2024can}.
}
\label{fig:interpretability}
\vspace{-5mm}
}
\end{figure*}

\subsection{Cross-lingual embeddings}

Many features of natural language, such as parts-of-speech, intensifiers, and modifiers, recur across languages.
For example, while English and Sanskrit have different inflections and character sets, they exhibit similar verb conjugations and noun declensions.
Machine translation models must disentangle these distinctions to construct maps between different languages' embedding spaces. 
A common approach is an encoder-decoder translation model, which trains a model to map sentences onto a universal representation in a latent space. For example, a Sanskrit encoder maps a sentence into the latent space, and an English decoder then translates it \cite{wu2016google}. Syntax information, such as word ordering or inflection, is typically distinct among languages and thus not necessarily preserved in the latent space. In contrast, manifolds associated with semantic content remain conserved, and the low-dimensional latent coordinates capture information such as token positions and parts-of-speech \cite{artetxe2019massively,chang2022geometry}. 
Taking this approach even further, cross-lingual translation models construct a single shared latent space from many languages. These models typically outperform single-pair translation models, particularly for languages with less available training data, like Swahili or Urdu \cite{conneau2019unsupervised}.

Could the same effect hold for rare cell types? One analogy for cross-lingual latent spaces is shared embeddings of cell types across distinct organisms. Statistical alignment methods may be used to combine cell type populations across species with similar tissue groups (Fig. \ref{fig:interpretability}A) \cite{butler2018integrating,stein2019decomposing,song2023benchmarking,tarashansky2021mapping,kriebel2022uinmf,yang2024genecompass,pearce2025cross}. Like understudied languages, rarer cell types benefit from integrated analysis; for example, in a joint embedding of human and mouse pancreatic cells, a combined embedding better resolves subpopulations associated with stress response during protein assembly \cite{butler2018integrating}.
Recent works extend this concept by proposing universal cell embeddings, in which a single foundation model is trained on data spanning subjects, species, and even sequencing modalities \cite{rosen2023universal,lopez2018deep,rosen2024toward}. The resulting embedding exhibits emergent properties, including zero-shot embedding of new species or tissues without retraining.

As single-cell foundation models become more contextual and high-dimensional, the geometry of their embeddings may encode nontrivial biological structure, such as indirect regulatory grammars.
One approach to probing embedding geometry is topological data analysis, a set of tools for analyzing high-dimensional point clouds.
In language models, these methods detect cusp-like singularities that form due to polysemy \cite{jakubowski2020topology}.
On gene expression data, topological methods quantify the degree to which processes like developmental lineages produce low-dimensional manifolds or branches \cite{palande2023topological,korem2015geometry}. 
Newly-introduced robust statistical estimators of the intrinsic dimensionality of point clouds may be used to probe internal representations in artificial neural networks (Fig. \ref{fig:interpretability}B, left) \cite{facco2017estimating}. 
Recent results use these estimators to show that the intrinsic dimensionality of gene expression correlates with pluripotency, across diverse taxa ranging from mice to zebrafish \cite{biondo2024intrinsic}.
Several approaches directly constrain representations to enforce particular topological features. For example, imposing hyperbolic structure on embeddings improves resolution of branching processes associated with differentiation \cite{kuang2025reconstructing,schluter2025integrating,klimovskaia2020poincare,zhou2021hyperbolic,ding2021deep,bhasker2025uncovering}. 

The emerging field of \textit{mechanistic interpretability} examines the reasoning and internal representations of large language models.
One such approach, linear probes, trains small linear regression models to predict particular linguistic features, like parts-of-speech or subject-verb agreement, from latent states or internal activations of layers \cite{alain2016understanding,mamou2020emergence,belinkov2019analysis}. 
This approach quantifies how explicitly different features are represented, and can identify where semantic versus grammatical information resides within the model.
In single-cell foundation models, linear probes identify gene families that the model weighs particularly highly in making predictions, such as by highlighting inflammation and heatshock genes in immune cell datasets \cite{pedrocchi2024identifying}
However, linear probes typically require a supervision signal, such as a ground truth dataset showing known effects of knockdowns, motivating the need for unsupervised methods. Sparse autoencoders train shallow, wide neural networks to encode the activations of individual layers of large models (Fig. \ref{fig:interpretability}B, right). The width of the latent space, coupled with a strong sparsity penalty, encourages sparse autoencoders to map single concepts onto each latent dimension, thus unfolding polysemous activations in the original large model \cite{gao2024scaling}. 
In single-cell foundation models, sparse autoencoders isolate cell types that otherwise would be difficult to distinguish in embeddings \cite{schuster2024can}.

\subsection{Task-independence and amortization of reasoning}

Word embeddings derive their utility from their independence from downstream tasks. Training frontier models typically requires access to large amounts of computing resources, with contemporary models like RoBERTa-large optimizing as many as $10^{9}$ parameters over $10^{12}$ language tokens \cite{liu2019roberta}.
However, once trained, these models may be used as a preprocessing step for downstream tasks like sentiment classification. Single-cell technologies share a goal of identifying general representations that foreground relevant biological variables, while removing uninformative variation like batch or technical effects \cite{rosen2023universal}. Embeddings thus represent one motivation for the emerging foundation model paradigm in both language modelling and single-cell analysis, which argues that large-scale pretraining on diverse datasets leads to simpler starting representations for smaller-scale tasks. Task-independent embeddings thus serve to amortize computation. 

Large-scale pretrained models exhibit \textit{inference-time computation}, in which they spontaneously solve new tasks without additional training \cite{kojima2022large}. 
For example, large language models can be prompted to produce poetry with meter and scansion that are unseen in their training corpus \cite{walsh2024does}.
The underlying mechanism, \textit{in-context learning}, exploits the emergent ability of large models not only to retrieve, but also process, information during inference. 
Inference-time symbolic reasoning appears to improve with model scale, with language models recently advancing from solving elementary-school word problems to standardized mathematics exams for undergraduates  \cite{wu2024inference,liu2024mathbench,ahn2024large}.
Achieving similar results for biological datasets represents a frontier for single-cell foundation models.
Several recent models exhibit forms of inference-time reasoning, such as zero-shot embedding of novel cell types, prediction of protein interactions, and anticipation of responses to genetic perturbations \cite{cui2024scgpt,theodoris2023transfer}.
However, these tasks have unclear difficulty compared to language modeling tasks like standardized exams, leading to conflicting results regarding the efficacy of current single-cell foundation models \cite{kedzierska2023assessing,csendes2025benchmarking,ahlmann2025deep,wenteler2024perteval}. 
A better test may be the ability of large models to decipher the indirect, multiscale, and highly nonlinear logic of many regulatory circuits.
For example, the immune system implements elaborate combinatorial receptor-ligand interactions, phosphorylation cascades, and feedback loops, in order to discriminate self from non-self antigens \cite{germain2001art,chakraborty2014insights}. Parsing these logical circuits is akin to solving a complex mathematical reasoning problem, requiring models that can effectively process symbolic information.

\subsection{Conclusion \& Limitations of the analogy}

Drawing parallels between language models and single-cell embeddings reveals shared principles in how high-dimensional spaces encode structured, context-dependent information. However, the analogy between cells and word tokens has natural limits, presaging potential limitations of foundation models for single-cell genomics.

In natural languages, "context" arises from discrete, ordered sequences, where exact position and co-occurrence of tokens convey meaning \cite{harris1954distributional,firth1957synopsis}. 
In contrast, a cell's context arises from a web of spatial relationships, signaling interactions, lineage history, and environmental conditions---most of which are not naturally represented as ordered sequences. 
No two cells are exact replicates, and their surrounding biochemical and environmental context can never be fully reproduced \cite{stuart2019integrative,hicks2018missing}. Furthermore, unlike words in a corpus, cells cannot be resampled from the same underlying distribution without perturbing the system, limiting the robustness and stationarity assumptions of statistical analogies.
A key challenge for emerging virtual cell models will thus be their ability to distill informative context in order to resolve polysemy in cell states while still finding concise representations. 
Other challenges include integrating diverse experimental contexts without loss of biological specificity, and capturing nonlinear regulatory logic within embedding spaces \cite{fang2024scmae,eraslan2019single,fu2025foundation,cui2024scgpt}. 

In contrast to language, where token context is explicit and uniformly structured, biological context is multiscale, incomplete, and often indirect.
Moreover, while neighboring words directly inform language token context, a cell's relevant "neighbors" may be defined in multiple, potentially conflicting ways (physical proximity, developmental stage, functional similarity). 
Resolving this ambiguity requires contextual cell embeddings, integrating heterogeneous modalities such as spatial transcriptomics, proteomics, chromatin accessibility, or lineage tracing to derive a unified representation.
Truly multimodal foundation models offer a potential solution, by treating auxiliary information---like gene ontologies, medical literature, or known regulatory hierarchies---on an even footing with expression data, thus decoupling modality-specific factors from informative biological variation \cite{hu2025regformer,levine2024cell2sentence,rizvi2025scaling,theodoris2023transfer}.
However, combining modalities at scale raises technical challenges: batch effects, inconsistent coverage among different modalities, and the difficulty of defining context windows across different samples \cite{armingol2021deciphering,bastidas2019comprehensive}. 
Even if relevant auxiliary information is available, its incorporation into embeddings can amplify biases in the experimental design, leading to overfitting to specific tissue types, organisms, or experimental protocols \cite{rosen2023universal,cui2024scgpt}.
Identifying such effects will be necessary in future virtual cell models, and represents an area where mechanistic interpretability and low-dimensional manifold discovery may prove informative.

The cell token analogy also breaks down when considering the dynamical nature of biological systems. In languages, dynamic embeddings model variability in token meaning without altering the underlying corpus. Yet in biology, a cell's "meaning" irreversibly changes over time through differentiation, signaling, senescence, and adaptation \cite{la2018rna,haghverdi2016diffusion,bergen2020generalizing}. 
Capturing these processes requires embedding models that are temporally aware, capable of representing continuous trajectories, and robust to sparse or noisy longitudinal data. 
Moreover, truly contextual embeddings for cells must incorporate causal relationships---distinguishing between correlation and regulatory influence---a level of mechanistic grounding without an obvious equivalence in grammatical rules. 
Improved benchmarks, which test the ability of foundation models to parse complex and indirect regulatory logic, will help help transform cell representations from descriptive maps into predictive, reasoning-ready representations for biology.

\section{Acknowledgments}
W.G. thanks Ski Krieger and the members of the Chan-Zuckerberg theory in biology group. W.G. thanks Medici Therapeutics.This project has been made possible in part by Grant No. DAF2023-329596 from the Chan Zuckerberg Initiative DAF, an advised fund of Silicon Valley Community Foundation. W.G. is supported by NSF DMS 2436233 and NSF CMMI 2440490. The funders had no role in study design, data collection and analysis, decision to publish, or preparation of the manuscript.

\bibliography{cites}

\end{document}